\pdfoutput=1

\documentclass[11pt,twoside,a4paper,cmspaper,final,collab]{cms-tdr}
\def\svnVersion{240278}\def\cmsCernNo{2013-208}\def\cmsCernDate{\today}\def\cmsMessage{Published in Physical Review Letters as \href{http://dx.doi.org/10.1103/PhysRevLett.11.171802}{\doi{10.1103/PhysRevLett.11.171802.}}}

\begin{document}\cmsNoteHeader{TOP-12-037}

\hyphenation{had-ron-i-za-tion}
\hyphenation{cal-or-i-me-ter}
\hyphenation{de-vices}

\RCS$Revision: 234748 $
\RCS$HeadURL: svn+ssh://svn.cern.ch/reps/tdr2/papers/TOP-12-037/trunk/TOP-12-037.tex $
\RCS$Id: TOP-12-037.tex 234748 2014-04-01 17:53:57Z alverson $
\newlength\cmsFigWidth
\ifthenelse{\boolean{cms@external}}{\setlength\cmsFigWidth{0.85\columnwidth}}{\setlength\cmsFigWidth{0.4\textwidth}}
\ifthenelse{\boolean{cms@external}}{\providecommand{\cmsLeft}{top}}{\providecommand{\cmsLeft}{left}}
\ifthenelse{\boolean{cms@external}}{\providecommand{\cmsRight}{bottom}}{\providecommand{\cmsRight}{right}}
\ifthenelse{\boolean{cms@external}}{\providecommand{\NA}{\ensuremath{\cdots}}}{\providecommand{\NA}{\text{---}}}
\providecommand{\mt}{\ensuremath{m_\mathrm{T}}\xspace}
\providecommand{\met}{\ETslash}
\providecommand{\st}{\ensuremath{\mathrm{S}_\mathrm{T}}\xspace}
\providecommand{\rj}{\ensuremath{\mathrm{j}}}
\cmsNoteHeader{TOP-12-037} 
\title{\texorpdfstring{Search for flavor-changing neutral currents in top-quark decays
$\cPqt \to \cPZ\cPq$ in $\Pp\Pp$ collisions at $\sqrt{s} = 8$\TeV}{Search for flavor-changing neutral currents in top-quark decays t to Zq in pp collisions at sqrt(s)=8 TeV}}

\date{\today}

\abstract{
A search for flavor-changing neutral currents in top-quark decays
$\cPqt\to \cPZ\cPq$ is performed in events produced from the
decay chain $\cPqt \cPaqt\to \cPZ\cPq+\PW\cPqb$, where
both vector bosons decay leptonically, producing a final state with
three leptons (electrons or muons).
A data set collected with the CMS detector at the LHC is used,
corresponding to an integrated luminosity of 19.7\fbinv of
proton-proton collisions at a center-of-mass energy of 8\TeV.
No excess is seen in the observed number of events relative to the
standard model prediction; thus no evidence for flavor-changing neutral
currents in top-quark decays is found.
A combination with a previous search at 7\TeV excludes a $\cPqt \to \cPZ\cPq$
branching fraction greater than 0.05\% at the 95\% confidence level.
}

\hypersetup{%
pdfauthor={CMS Collaboration},%
pdftitle={Search for flavor-changing neutral currents in top-quark decays
t to Zq in pp collisions at sqrt(s)=8 TeV},%
pdfsubject={top},%
pdfkeywords={CMS, physics, top, searches}}

\maketitle 

\label{sec:introduction}
\par

The heaviest known elementary particle, the top quark, decays
to a bottom quark and a $\PW$ boson, $\cPqt \to \PW\cPqb$, with a branching
fraction of nearly 100\%~\cite{Beringer:1900zz}.
Within the standard model (SM),
the corresponding flavor-changing neutral current (FCNC) decay to a
$\cPZ$ boson and a
light up-type quark ($\cPqu$ or $\cPqc$), $\cPqt \to \cPZ\cPq$,
is suppressed by the GIM mechanism~\cite{ref:GIM},
occurring only at the quantum loop level, with a branching  fraction
$\mathcal{B}(\cPqt \to \cPZ\cPq)$ at
$\mathcal{O}(10^{-14})$~\cite{ref:Glover:2004cy}. The detection of
FCNC $\cPqt \to \cPZ\cPq$ decays at a higher-than-expected rate
would thus be clear evidence for physics beyond the SM.

Some extensions of the SM, such as $R$-parity-violating supersymmetric
models~\cite{Yang:1997dk}, top-color-assisted technicolor
models~\cite{Lu:2003yr}, and singlet quark models~\cite{AguilarSaavedra:2002kr}
predict enhancements of the FCNC branching fraction that
could be as large as $\mathcal{O}(10^{-4})$.
These models, however, need to be updated from their earlier parametrizations
using the latest Large Hadron Collider (LHC) results. While this is not easily
done in general without
a detailed analysis, a recent study~\cite{Agashe:2013hma}
places $\mathcal{B}(\cPqt \to \cPZ\cPq)$ at $\mathcal{O}(10^{-5})$
in warped extra dimension models~\cite{Agashe:2004ay, Agashe:2006wa}.
The rate is very sensitive to the Kaluza-Klein gluon scale $m_{\rm KK}$,
as well as right-handed mixing parameters.
The $m_\mathrm{KK}$ scale is probed
directly~\cite{Aad:2012dpa, Chatrchyan:2013lca}
at the LHC, 
while 
B physics measurements~\cite{Fox:2007in}
do not significantly constrain right-handed couplings.
In this Letter we report
a search for $\cPqt \to \cPZ\cPq$ at the LHC,
with results that start to complement
both the direct search for KK gluons,
as well as flavor physics constraints.

The experimental searches on top FCNC have been carried out
since LEP and HERA on single top quark production with best limit
$\mathcal{B}(\cPqt \to \cPZ\cPq) < 4\%$~\cite{Abdallah:2003wf,
Abramowicz:2011tv}.
The current limit by Tevatron on top FCNC decay is
${<}3.2\%$~\cite{ref:Abazov:2011qf}.
In a previous search with the Compact Muon Solenoid (CMS) experiment,
we reported results from a
search for this decay using 5.0\fbinv of proton-proton collisions at
a center-of-mass energy $\sqrt{s} = 7$\TeV,
resulting in a 95\% confidence level (CL)
upper limit on $\mathcal{B}(\cPqt \to \cPZ\cPq)$ of
0.21\%~\cite{Chatrchyan:2012hqa}. 
A limit of 0.73\%~\cite{ref:ATLAS-2012} on the branching fraction
has also been reported by the ATLAS
experiment from an analysis of 2.1\fbinv of 7\TeV data.
The analysis described in this letter uses a data sample corresponding to an
integrated luminosity of 19.7\fbinv of pp collisions at $\sqrt{s} = 8$\TeV.

The central feature of the CMS
apparatus is a superconducting solenoid, which provides an axial magnetic
field of 3.8\unit{T}. Within the field volume there are a silicon pixel and
strip tracker, a lead tungstate crystal electromagnetic calorimeter, and a
brass/scintillator hadron calorimeter. Charged-particle trajectories are
measured by
the tracker, covering
$0\le \phi \le 2\pi$ in azimuth and $\abs{\eta} < 2.5$ in pseudorapidity,
where $\eta$ is defined as $-\ln[\tan (\theta/2)]$ and $\theta$ is the polar
angle of the trajectory of the particle with respect to the counterclockwise
proton beam direction. Muons are identified and measured in gas-ionization
detectors embedded in the steel flux return yoke outside the solenoid. A more
detailed description of the CMS detector can be found
in Ref.~\cite{Chatrchyan:2008zzk}.

Monte Carlo (MC) samples of Drell--Yan events,
SM $\ttbar$,
$\cPZ \ttbar$, $\PW \ttbar$, $\cPqt \cPqb\cPZ$,
and diboson ($\PW\PW$, $\PW\cPZ$, and $\cPZ\cPZ$) events are simulated using
$\MADGRAPH$~\cite{ref:madg},
while single-top-quark events are generated
using \POWHEG~\cite{ref:Nason:2004rx,ref:Frixione:2007vw,ref:Alioli:2010xd}.
The signal sample $\Pp\Pp \to \ttbar \to \cPZ\cPq + \PW\cPqb \to$
$\ell^+\ell^- \cPq + \ell'^\pm\nu\cPqb\ ( \ell,\, \ell' = \Pe , \Pgm, \tau)$ is
generated with \MADGRAPH.
One of the top quarks of the pair is forced to decay as $\cPqt \to \cPZ\cPq$,
where $\cPq$ stands for a $\cPqu$ or $\cPqc$ quark with equal probability,
while the other decays to $\PW \cPqb$.
The ratio between the dimension-4 vector and the axial-vector couplings of the
$\cPqt \to \cPZ\cPq$ FCNC model is assumed to be SM-like~\cite{Han:1998yr}.
In all cases, hadronization and showering are simulated with
\PYTHIA 6~\cite{ref:pythia}, while
$\tau$ decays are simulated with \TAUOLA~\cite{Was:2000st}.
The set of parton distribution functions (PDFs) used is
CTEQ6L~\cite{PhysRevD.78.013004}.
The CMS detector response for all MC samples is simulated using a
$\GEANTfour$-based~\cite{Agostinelli2003250} model, and the events are
reconstructed and analyzed using the same software used to process
collision data. The simulated events are weighted so that the trigger
efficiencies, reconstruction efficiencies, and the distribution
of reconstructed vertices observed in data are reproduced.

\label{sec:preselection}

The search is performed by looking for $\cPqt\overline\cPqt$ events where one
top quark decays into $\cPZ\cPq$ and the other decays into $\PW\cPqb$
with both vector bosons decaying leptonically, which provide a very clear
signature. The analysis follows closely the search performed at
7\TeV~\cite{Chatrchyan:2012hqa}.
Several of the event selection requirements have been re-optimized
before the complete dataset was collected,
based on the expected signal and background yields at 8\TeV
with $\mathcal{B}(\cPqt\to \cPZ\cPq) = 0.1\%$.

Events are required to pass at least one of the $\Pe\Pe$ or $\mu\mu$
high transverse momenta ($\pt$) dilepton triggers. Events with two
opposite-sign, same-flavor, isolated leptons ($\Pe$ or $\mu$) having
an invariant mass between 78\GeV and 102\GeV, consistent with a $\cPZ$-boson
decay, and one extra charged lepton ($\Pe$ or $\mu$) are selected.
When there is more than one lepton pair forming a $\cPZ$ candidate, the pair
with invariant mass closest to the nominal value is taken.
All three leptons must satisfy
the following kinematic requirements:
$\pt>20$\GeV and $\abs{\eta} < 2.5$ for electrons and
$\abs{\eta} < 2.4$ for muons.
The lepton selection efficiencies (reconstruction, identification, and
isolation) mean values and their dependence with \pt and $\abs{\eta}$ are
consistent between the data and the
simulation~\cite{Chatrchyan:2011cm, Chatrchyan:2012bra}.

Multiple simultaneous interactions per bunch crossing (pileup) were observed
in data. Events are required to have at least one good primary vertex
candidate~\cite{CMS-PAS-TRK-10-005}.
In events with more than one candidate, the vertex with highest
$\Sigma {\pt}^2$ of its associated tracks is selected. The leptons and all
charged particle tracks that are associated with jets are required to be
consistent with originating from the primary vertex.

Since the leptons are expected to originate from the decays of $\PW$ and
$\cPZ$ bosons, they are required to be isolated as defined in
Ref.~\cite{Chatrchyan:2012xi}.
Events with a fourth isolated lepton are rejected.
Neutrinos from $\PW$-boson decays escape detection and produce a significant
momentum imbalance in the detector in the plane transverse to the beams.
The missing transverse momentum
($-\Sigma \ptvec$) and its magnitude ($\ETslash$)
are reconstructed using the CMS particle-flow technique~\cite{ref:pf}, and
we require the \ETslash to be larger than 30\GeV.
The $\PW$-boson candidates are constructed from the momentum of the extra
lepton and the missing transverse momentum (assumed to originate from
an undetected neutrino), by constraining the resulting invariant mass to be
equal to the $\PW$-boson mass~\cite{Beringer:1900zz}.

The requirements described above, namely events with dilepton-triggers,
a $\cPZ$ boson candidate, an extra lepton, no fourth lepton, and the
requirement of \ETslash, will be referred to as the ``basic event selection''.
The observed number of events after the basic event selection
is 1424, 
in agreement with the MC expectation of $1455\pm16$ events, including
$1229\pm4$ events from $\PW\cPZ$ and $86.3\pm0.2$ events from $\cPZ\cPZ$
production, where the uncertainty quoted is statistical only.

\begin{figure*}[htb!]
\begin{center}
\includegraphics[width=0.32\textwidth]{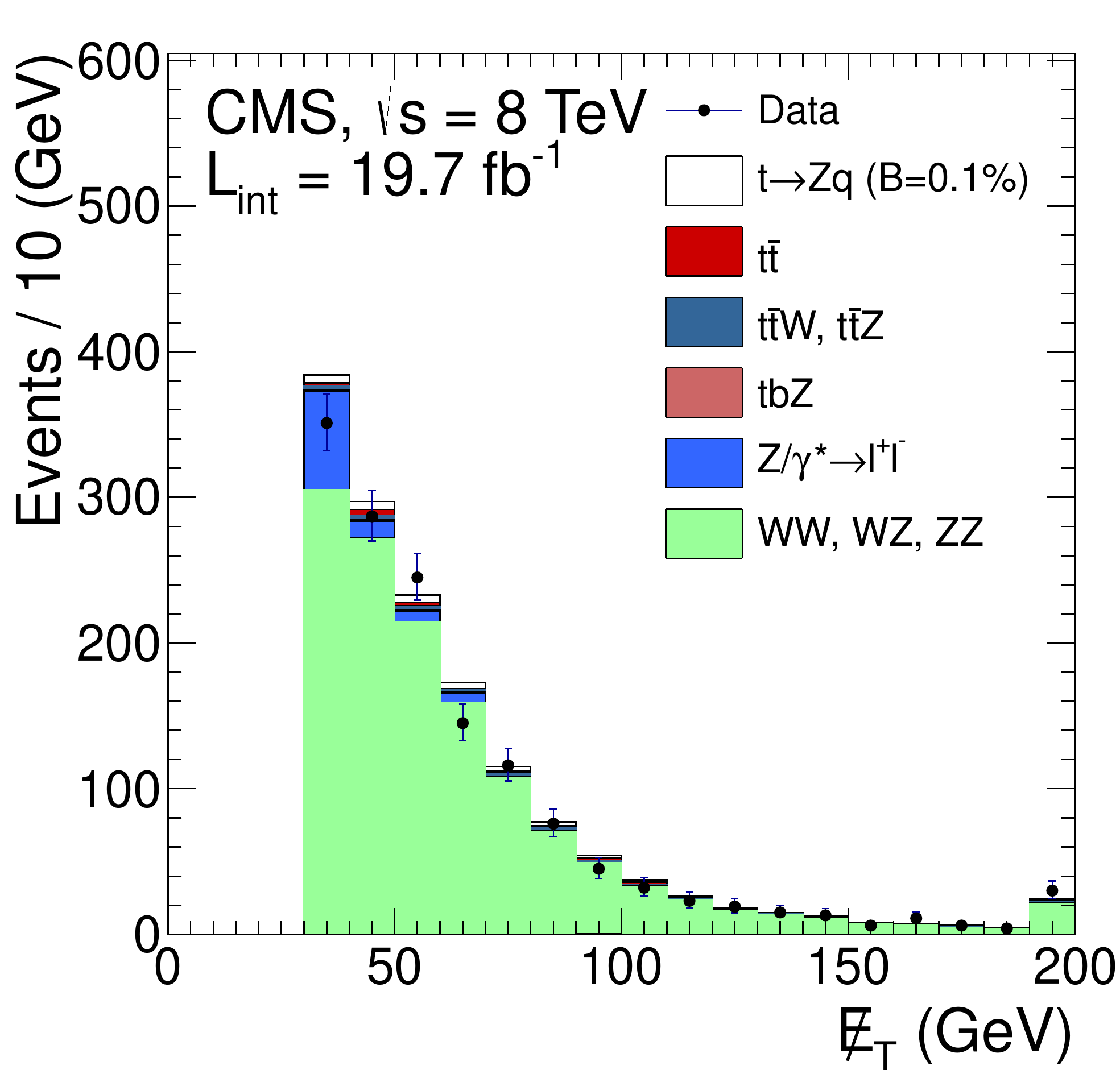}
\includegraphics[width=0.32\textwidth]{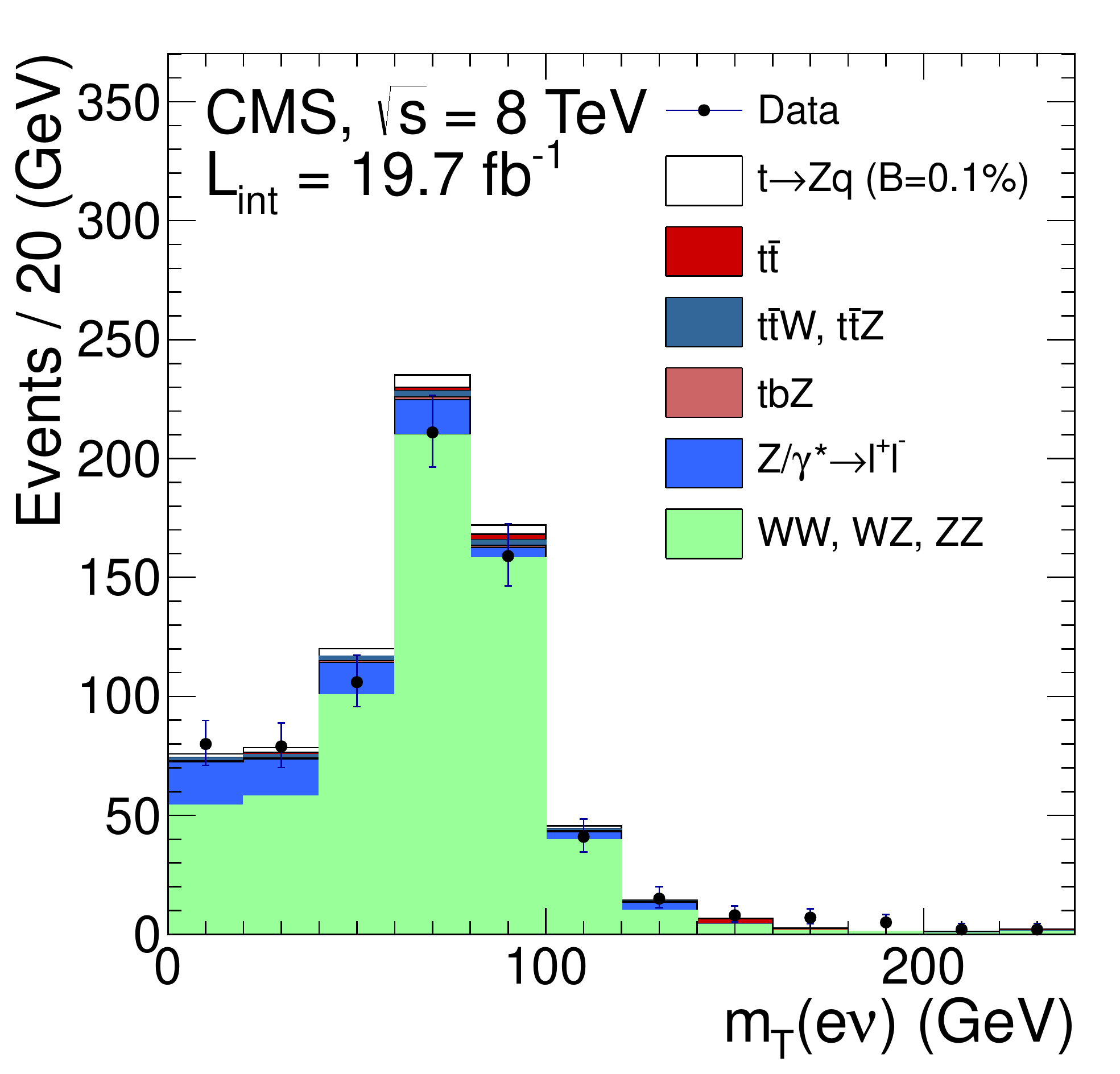}
\includegraphics[width=0.32\textwidth]{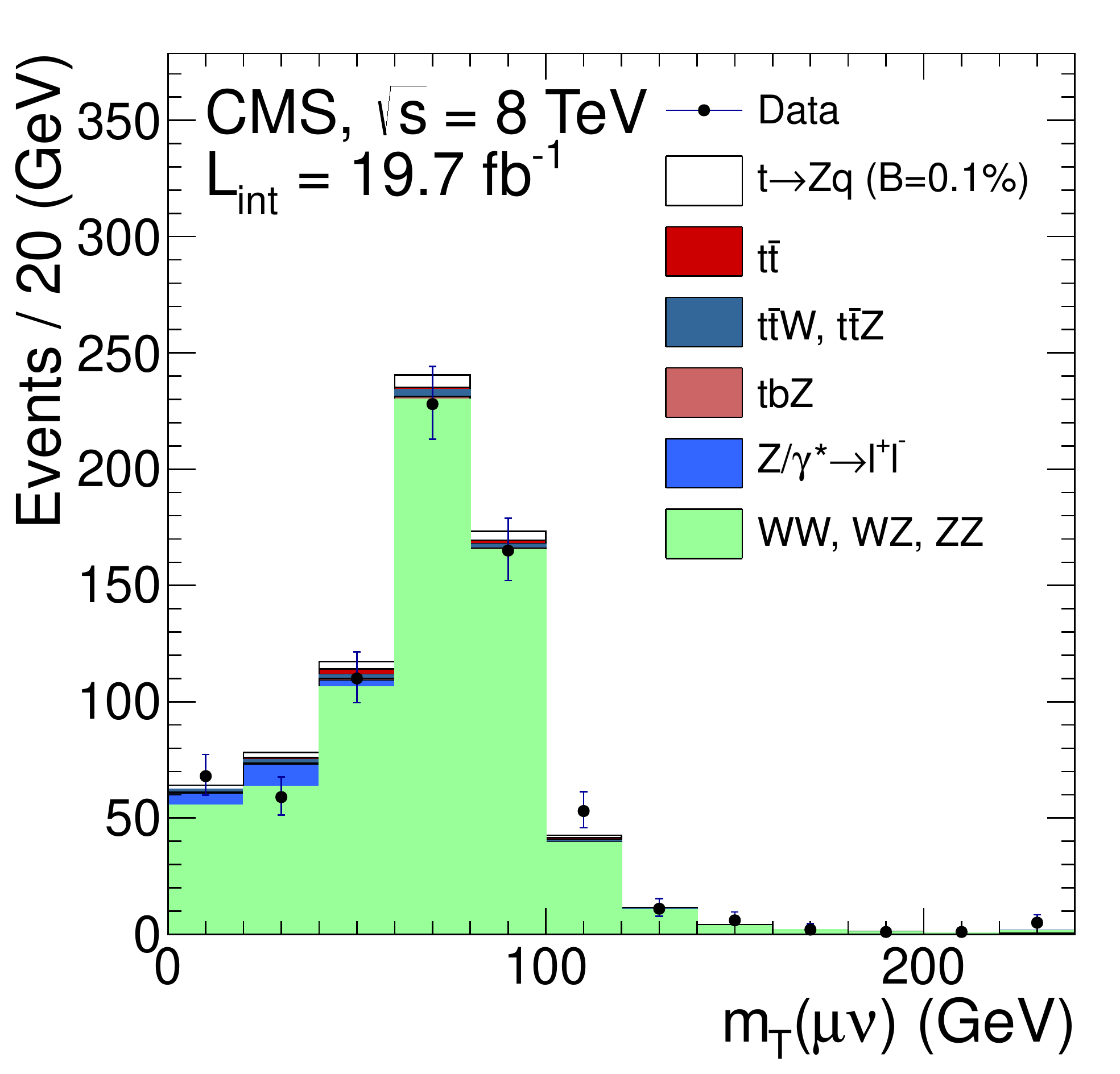}
\caption{\label{fig:jet} Comparison between data and simulated events
for an integrated luminosity of 19.7\fbinv, after the basic event selection,
for the \ETslash distribution (left), the reconstructed $\Pe \nu$ transverse
mass (middle) of the $\PW$-boson candidate, and the reconstructed $\mu \nu$
transverse mass (right) of the $\PW$-boson candidate.
The data are represented by the points with error bars,
and the open histogram shows the expected signal
assuming $\mathcal{B}(\cPqt\to \cPZ\cPq)$ is equal to 0.1\%. Stacked solid
histograms represent the dominant backgrounds, with statistical uncertainties
on these backgrounds at the few percent level (not shown).
}
\end{center}
\end{figure*}

Figure~\ref{fig:jet} shows the distributions for data and simulated events of
the \ETslash and transverse mass (\mt) of
the $\PW$-boson candidate
after the basic event selection.
The transverse mass is calculated using the transverse momentum of the extra
lepton ($\pt^\ell$), the \ETslash, and the azimuthal angle
difference ($\Delta\phi$) between the two, as
$\mt=\sqrt{\smash[b]{2 \pt^\ell\met(1-\cos(\Delta\phi))}}$.

\label{sec:signal}

To reduce the background from diboson events we require at least two jets,
reconstructed also using a particle-flow technique~\cite{ref:pf},
each with $\pt > 30$\GeV and $\abs{\eta} < 2.4$.
Exactly one of these jets should be identified (tagged) as a $\cPqb$-quark jet.
These requirements further reduce the observed event yields from 1424 after
the basic event selection to 29.
The $\cPqb$-jet identification is performed using the combined
secondary vertex $\cPqb$-tagging algorithm 
described
in Ref.~\cite{Chatrchyan:2012jua}.
This tagging method has an identification efficiency of
$62\%$ for $\cPqb$ jets with transverse momentum between 30\GeV and
100\GeV and a misidentification rate of about 18\% for $\cPqc$ jets
and below 1.5\% for other jets.

\begin{figure*}[htb!]
\begin{center}
\includegraphics[width=0.32\textwidth]{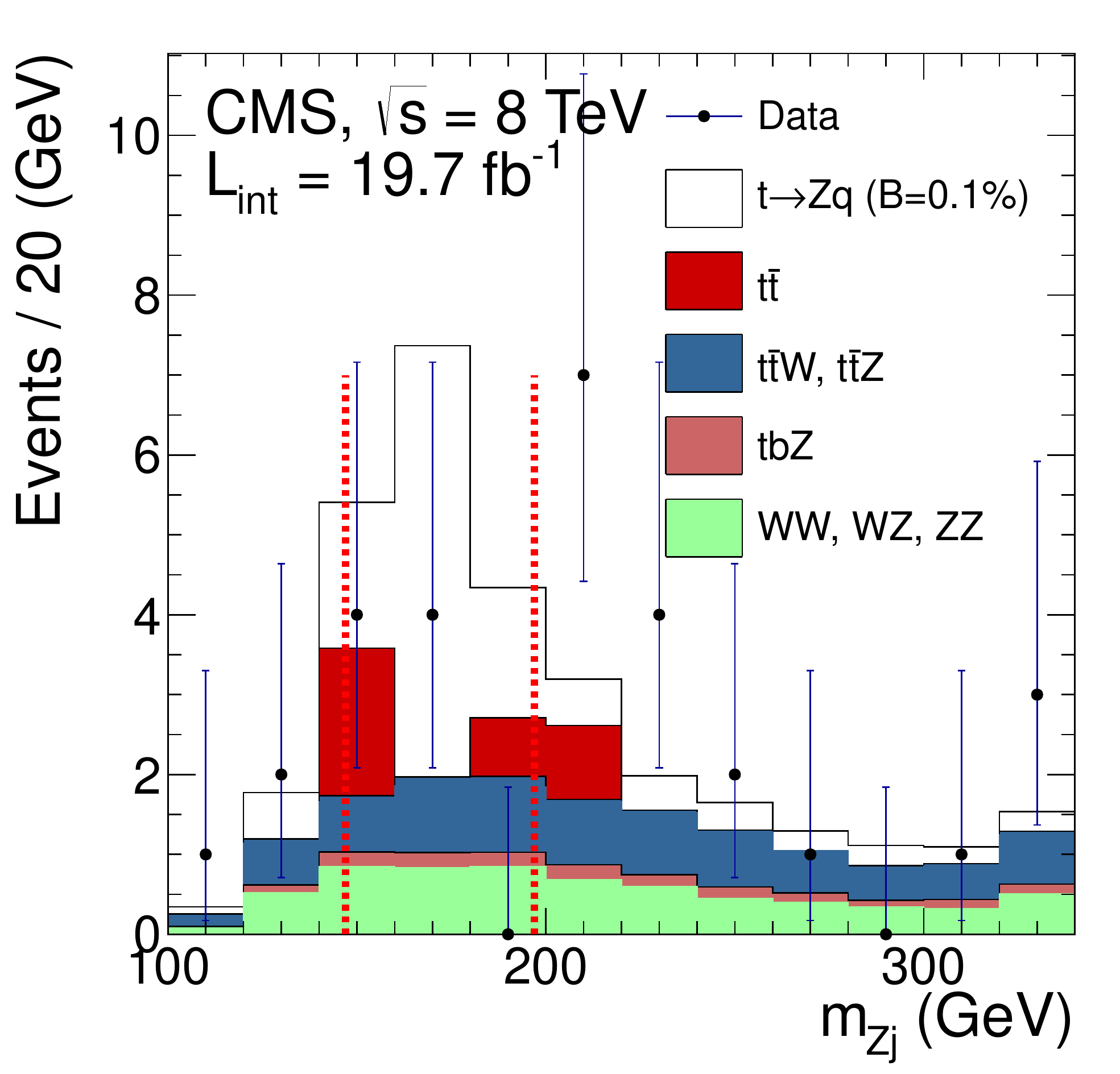}
\includegraphics[width=0.32\textwidth]{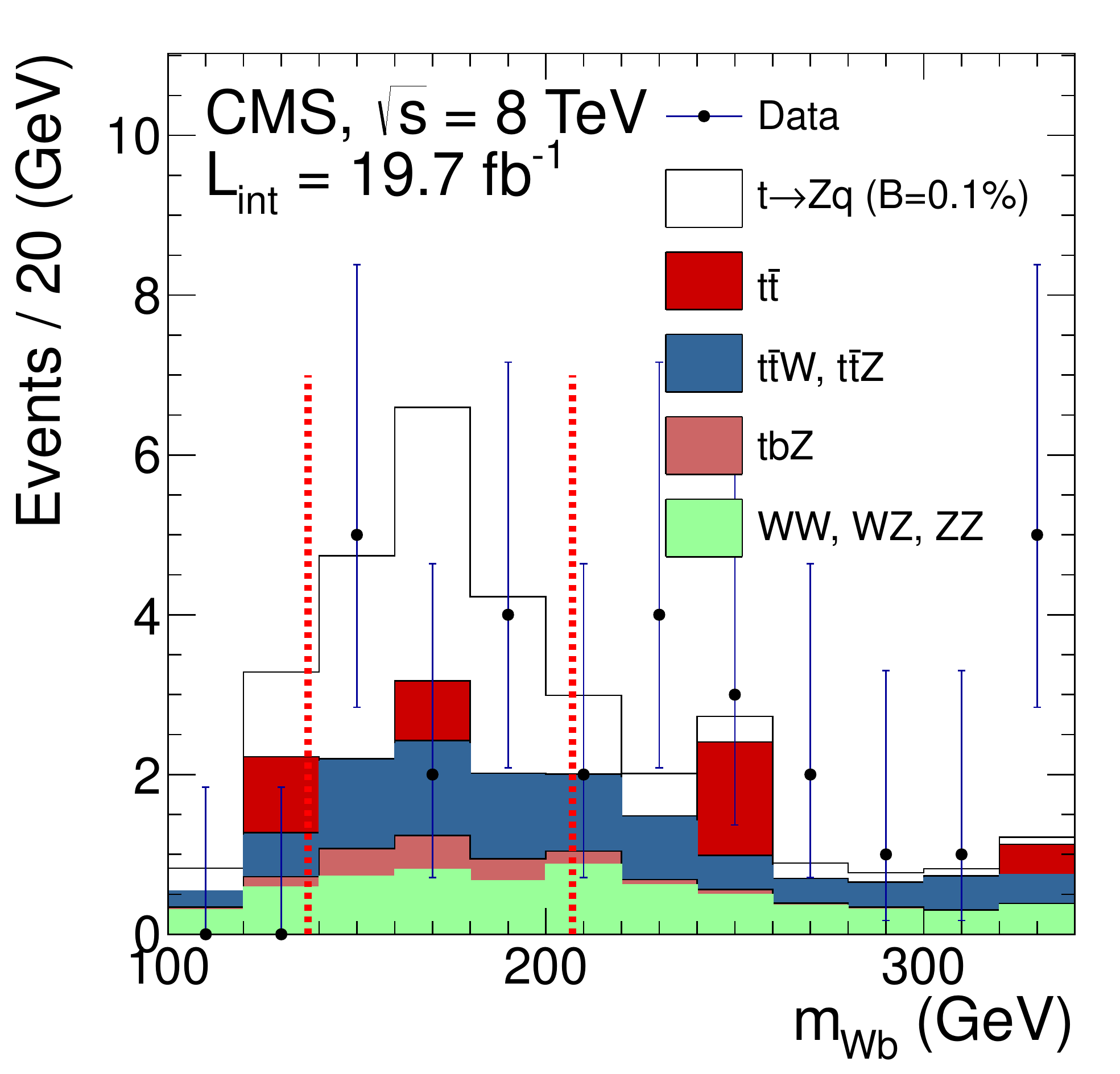}
\includegraphics[width=0.32\textwidth]{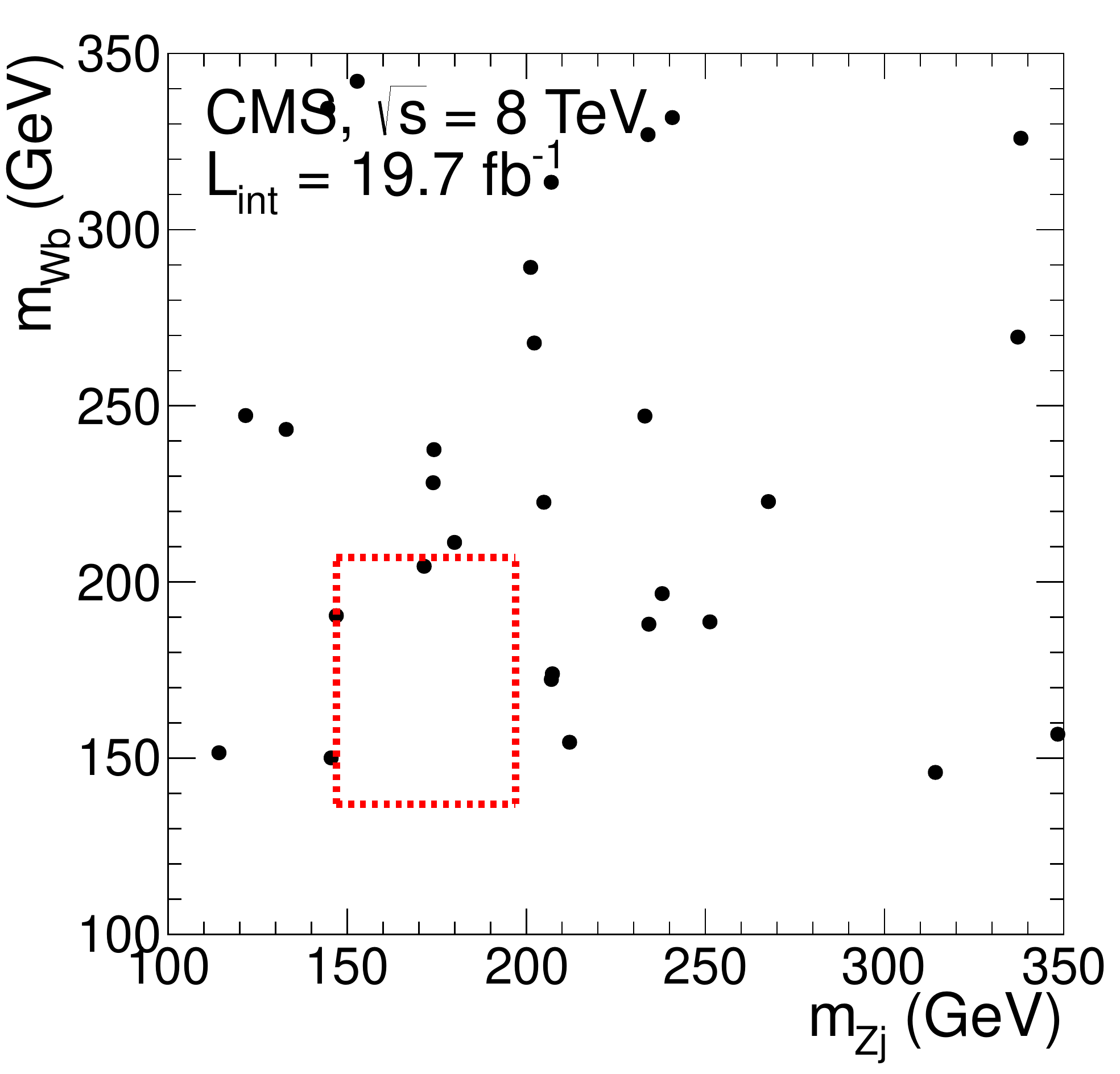}
\caption{\label{fig:mass}
Comparison between data and simulated events of the $m_{\cPZ \rj}$ (left),
$m_{\PW\cPqb}$ (middle), and 2D scatter (right) distributions
after the event selection prior to the top-quark mass requirements, which are
shown as the dotted vertical lines (left, middle) and box (right). The data,
corresponding to an integrated luminosity of 19.7\fbinv, are represented
by the points with error bars and the open histogram is the expected signal.
The stacked solid
histograms represent the dominant backgrounds.
The statistical uncertainties
are not drawn.
The last bin in each of the left two plots contains
all the overflow events.
}
\end{center}
\end{figure*}

The invariant mass of the $\PW$ boson and the $\cPqb$-tagged jet,
$m_{\PW\cPqb}$, is required to be within 35\GeV
of the top-quark mass,
which is set to 172.5\GeV in the simulation. A non-$\cPqb$
jet is combined with the $\cPZ$ candidate to form a second top-quark candidate. By
examining all possible pairings, the top-quark candidate which has the largest
separation in azimuthal angle to the first top quark is selected, and the
reconstructed top-quark mass, $m_{\cPZ \rj}$, is required to be within 25\GeV
of the assumed value of 172.5\GeV. The mass requirements 
are the same as in the 7\TeV analysis~\cite{Chatrchyan:2012hqa}.
Figure~\ref{fig:mass}\, shows the comparison of the $m_{\cPZ \rj}$ and
$m_{\PW\cPqb}$ distributions in data and simulation, while Table~\ref{tab:effy}
summarizes the 
signal efficiencies determined from simulated events.

\begin{table}[htb]
\centering
\topcaption{
\label{tab:effy}
Signal selection efficiency, in percent, for each dilepton channels
with respect to events from all channels.
The efficiency is calculated as the fraction of events
with leptonically (${\Pe,}\,\mu,\,\tau$) decaying
$\PW$ and $\cPZ$ bosons passing the selection.
Only MC statistical uncertainties are shown.
}
\begin{scotch}{ c   c }
Channel & Efficiencies $[\%]$   \\
 \hline
$\Pe\Pe\Pe$    & 0.49 $\pm$ 0.01 \\
$\Pe\Pe\mu$    & 0.52 $\pm$ 0.01 \\
$\mu\mu \Pe$   & 0.55 $\pm$ 0.01 \\
$\mu\mu\mu$    & 0.55 $\pm$ 0.01 \\
\hline
All            & 2.12 $\pm$ 0.03 \\
\end{scotch}
\end{table}

\label{sec:back}

According to simulations, the dominant backgrounds arise from diboson and
$\ttbar+ \mathrm{X}$ production.
These processes can be categorized into three groups
based on the number of $\cPqb$ quarks present: (a)~diboson and Drell--Yan events with nearly no $\cPqb$ quarks;
(b)~events from top-quark FCNC decay with only one $\cPqb$ quark;
(c)~$\ttbar$, $\cPqt\cPqb\cPZ$, $\ttbar\PW$, and $\ttbar\cPZ$ processes with
at least two $\cPqb$ quarks. Events passing the basic
event selection, with two jets to be paired with $\PW$ and $\cPZ$ bosons
are divided into three samples:
(a)~events with no $\cPqb$-tagged jets;
(b)~events with exactly one $\cPqb$-tagged jet; and
(c)~the rest of the events.
The numbers of events in those three samples
can be related to
the yields of the three groups
based on the $\cPqb$-tagging efficiencies for $\cPqb$ jets,
$\cPqc$ jets, or other jets, which are
measured using data.
The numbers of events in the three groups are then turned into
an estimate of the corresponding yields via a linear $3 \times 3$ system
of equations to be solved before the top-quark mass requirements. The
corresponding acceptances of the mass requirements are obtained from MC.
The overall contribution
from $\PW\cPZ$ plus $\cPZ\cPZ$ and Drell--Yan backgrounds is
estimated to be $1.4\pm0.1\stat\pm0.3\syst$ events.
The expected yield from $\ttbar\PW$,
$\ttbar\cPZ$, $\cPqt\cPqb\cPZ$, and $\ttbar$ backgrounds is
$1.7\pm0.8\stat\pm0.4\syst$ events.
The uncertainty of the
$\cPqb$-tagging efficiency, measured in control data samples, and the
uncertainty on the top-quark mass requirement, estimated with MC simulation,
contribute to the systematic uncertainty.
The estimated background yields are summarized in
Table~\ref{tab:yields} and show a good agreement with
those obtained from MC simulation.
The background estimations from data are used for the final results.

\begin{table}[htb]
\centering
\topcaption{
\label{tab:yields}
Expected number of signal $\cPqt\to\cPZ\cPq$ events,
background composition, and observed events
corresponding to an integrated luminosity of 19.7\fbinv
for all dilepton channels; background estimates included.
The uncertainties in the background
estimation include the statistical and systematic components shown
separately, in that order. 
}
\begin{scotch}{ l  c  c }
Process  & Estimation from data &  MC prediction\\
\hline
 $\cPqt\to\cPZ\cPq$ ($\mathcal{B}=0.1\%$) & \NA & $6.4\pm0.1\pm1.3$ \\
\hline
 $\PW\cPZ$ & \multirow{3}{*}{$1.4\pm0.1\pm0.3$} & $0.9\pm0.1\pm0.3$ \\
 $\cPZ\cPZ$ &  & $< 0.1$ \\
 Drell--Yan &  & $< 0.1$ \\
\hline
 $\ttbar$     & \multirow{4}{*}{$1.7\pm0.8\pm0.4$} & $0.7^{+1.1}_{-0.4}\pm1.2$ \\
 $\ttbar\cPZ$ &  & $1.1\pm0.1\pm0.8$ \\
 $\ttbar\PW$  &  & $0.1\pm0.1\pm0.1$ \\
 $\cPqt\cPqb\cPZ$ &  & $0.3\pm0.1\pm0.2$ \\
\hline
 Total background & $3.1\pm0.8\pm0.8$ & $3.2\pm1.2\pm1.5$ \\
 \hline
 Observed events       &  1 & \NA \\
\end{scotch}
\end{table}

\label{sec:sys}
To calculate the expected upper limits,
the systematic uncertainties from the dilepton trigger efficiency,
lepton selection efficiency~\cite{Chatrchyan:2011cm},
pileup modeling~\cite{Antchev:2011vs},
$\cPqb$-jet tagging efficiency~\cite{Chatrchyan:2012jua},
jet energy scale and missing transverse energy resolution~\cite{ref:jetscale}
are included,
with the $\cPqb$-jet tagging efficiency being the dominant one for the
background estimation.
Additionally, several sources of uncertainties in the signal yield
are evaluated: the choice of PDFs,
generator parameters, and uncertainty in the $\ttbar$ cross section.
The major contributions come from the PDFs and the
generator parameters of the signal MC simulation. The prescription given in
Ref.~\cite{ref:cteq} is used to determine the uncertainty from the
CTEQ66 PDF error sets~\cite{ref:CTEQ6LL}.
The uncertainty from the generator parametrization is evaluated using CMS
fast simulation~\cite{Rahmat:2012fs} samples with different top-quark mass
assumptions ($\pm$2\GeV), different parton-jet matching thresholds (30\GeV and
60\GeV), and different event renormalization and factorization scales
(varied between $1/4$ and $4\times$ from their nominal value).
In addition, there is a 2.6\% uncertainty on the luminosity
measurement~\cite{CMS-PAS-LUM-13-001}.
All these sources, summarized in Table~\ref{tab:sys}, are combined in
quadrature to give a 20\% relative uncertainty in the signal selection
acceptance.
The systematic uncertainties in the background estimation are listed with
the total background prediction given in Table~\ref{tab:yields}.

\begin{table}[htb]
\centering
\topcaption{
\label{tab:sys}
Summary of the systematic uncertainties
for the signal selection
acceptance.}
\begin{scotch}{ l  c }
Source     & Uncertainty  \% \\
   \hline
Renormalization/factorization scales & 12 \\
Parton distribution functions & 7 \\
$\ttbar$ cross section & 7 \\
Parton matching threshold & 6 \\
Lepton selection  &  6  \\
Trigger efficiency &  5 \\
$\cPqb$-tagging &  5 \\
Top-quark mass & 4 \\
Jet energy scale  & 4 \\
Missing transverse energy resolution  & 3\\
Pileup modeling  & 3  \\
\hline
Total & 20 \\
\end{scotch}
\end{table}

\label{sec:results}

After applying all the criteria and adding all four channels,
$3.1\pm 1.1$ events are expected from SM background processes and
1 event is observed in data.
A 95\% confidence level (CL) upper limit on
the branching fraction of the $\cPqt\to\cPZ\cPq$ decay is determined
using the modified frequentist approach (CL$_\mathrm{s}$
method~\cite{ref:Junk:1999kv,ref:Read:2002hq}).
A summary of the observed and predicted yields and limits is presented in
Tables~\ref{tab:yields} and~\ref{tab:limits}.
The observed and expected 95\% CL upper limits on the branching
fraction $\mathcal{B}(\cPqt \to \cPZ\cPq)$ are 0.06\% and 0.10\%, respectively.

These results are combined with
the statistically-independent results of our
previous search~\cite{Chatrchyan:2012hqa}.
The systematic uncertainties
on the signal efficiency estimation and the luminosity measurements
are taken as fully correlated. 
Since the background estimations are based on independent samples,
their systematic uncertainties are treated as uncorrelated, except for the
uncertainties on the top mass selection requirement due to the choice of PDF,
which are also taken as fully correlated.
The combination with the 7\TeV b-tag analysis~\cite{Chatrchyan:2012hqa}
gives a slightly lower expected limit and hence is chosen
as the reference result. The observed upper limit on
$\mathcal{B}(\cPqt\to\cPZ\cPq)$ is 0.05\%, with a median
expectation of 0.09\%, and with $1\sigma$ and $2\sigma$ ranges which are
0.06--0.13\% and
0.05--0.18\%, respectively. The derived limits and their uncertainties
are shown in Table~\ref{tab:limits}.

\begin{table}[htb]
\centering
\topcaption{
\label{tab:limits}
Upper limits at a 95\% CL for $\mathcal{B}(\cPqt\to\cPZ\cPq)$, as obtained using
the 8\TeV data with an integrated luminosity of 19.7\fbinv, and from the
combination with previous CMS 7\TeV (5.0\fbinv) data.
}
\begin{scotch}{ l  c  c }
$\mathcal{B}(\cPqt\to\cPZ\cPq)$  & 8\TeV &  7\TeV + 8\TeV \\
\hline
 Expected upper limit &  $<$0.10\% & $<$0.09\% \\
\hline
 Observed upper limit &  $<$0.06\% & $<$0.05\% \\
\hline
 $1\,\sigma$ boundary &  0.06--0.13\% & 0.06--0.13\% \\
 $2\,\sigma$ boundary &  0.05--0.20\% & 0.05--0.18\% \\
\end{scotch}
\end{table}

\label{sec:conclusions}
In summary, a search for FCNC events in top-quark decays
in $\ttbar$ events produced in proton-proton collisions at $\sqrt{s} = 8$\TeV
is presented.
A sample of events with three leptons ($\Pe$ or $\Pgm$) in the final state
and compatible with leptonic decays of a $\cPZ$ and $\PW$ boson is selected
from data recorded with the CMS detector and corresponding to an integrated
luminosity of 19.7\fbinv. No excess of events above the background is observed.
Combining this result with a previous search corresponding to an integrated
luminosity of 5.0\fbinv at $\sqrt{s} = 7$\TeV, excludes a $\cPqt\to\cPZ\cPq$
branching fraction greater than 0.05\% at a confidence level of 95\%.
This new limit, about four times better than our previous one, has
been achieved with a better background estimation, increased cross section
at higher energy and increased integrated luminosity.
This new result starts to probe the enhanced FCNC branching fraction
predicted by certain SM extensions, such as warped extra
dimensions. Our limit can be translated into a constraint on
the KK gluon to be
heavier than 1.1 TeV~\cite{Agashe:2006wa}.

We congratulate our colleagues in the CERN accelerator departments for the excellent performance of the LHC and thank the technical and administrative staffs at CERN and at other CMS institutes for their contributions to the success of the CMS effort. In addition, we gratefully acknowledge the computing centers and personnel of the Worldwide LHC Computing Grid for delivering so effectively the computing infrastructure essential to our analyses. Finally, we acknowledge the enduring support for the construction and operation of the LHC and the CMS detector provided by the following funding agencies: BMWF and FWF (Austria); FNRS and FWO (Belgium); CNPq, CAPES, FAPERJ, and FAPESP (Brazil); MES (Bulgaria); CERN; CAS, MoST, and NSFC (China); COLCIENCIAS (Colombia); MSES and CSF (Croatia); RPF (Cyprus); MoER, SF0690030s09 and ERDF (Estonia); Academy of Finland, MEC, and HIP (Finland); CEA and CNRS/IN2P3 (France); BMBF, DFG, and HGF (Germany); GSRT (Greece); OTKA and NIH (Hungary); DAE and DST (India); IPM (Iran); SFI (Ireland); INFN (Italy); NRF and WCU (Republic of Korea); LAS (Lithuania); CINVESTAV, CONACYT, SEP, and UASLP-FAI (Mexico); MBIE (New Zealand); PAEC (Pakistan); MSHE and NSC (Poland); FCT (Portugal); JINR (Dubna); MON, RosAtom, RAS and RFBR (Russia); MESTD (Serbia); SEIDI and CPAN (Spain); Swiss Funding Agencies (Switzerland); NSC (Taipei); ThEPCenter, IPST, STAR and NSTDA (Thailand); TUBITAK and TAEK (Turkey); NASU (Ukraine); STFC (United Kingdom); DOE and NSF (USA).

\bibliography{auto_generated}   
\cleardoublepage \appendix\section{The CMS Collaboration \label{app:collab}}\begin{sloppypar}\hyphenpenalty=5000\widowpenalty=500\clubpenalty=5000\textbf{Yerevan Physics Institute,  Yerevan,  Armenia}\\*[0pt]
S.~Chatrchyan, V.~Khachatryan, A.M.~Sirunyan, A.~Tumasyan
\vskip\cmsinstskip
\textbf{Institut f\"{u}r Hochenergiephysik der OeAW,  Wien,  Austria}\\*[0pt]
W.~Adam, T.~Bergauer, M.~Dragicevic, J.~Er\"{o}, C.~Fabjan\cmsAuthorMark{1}, M.~Friedl, R.~Fr\"{u}hwirth\cmsAuthorMark{1}, V.M.~Ghete, C.~Hartl, N.~H\"{o}rmann, J.~Hrubec, M.~Jeitler\cmsAuthorMark{1}, W.~Kiesenhofer, V.~Kn\"{u}nz, M.~Krammer\cmsAuthorMark{1}, I.~Kr\"{a}tschmer, D.~Liko, I.~Mikulec, D.~Rabady\cmsAuthorMark{2}, B.~Rahbaran, H.~Rohringer, R.~Sch\"{o}fbeck, J.~Strauss, A.~Taurok, W.~Treberer-Treberspurg, W.~Waltenberger, C.-E.~Wulz\cmsAuthorMark{1}
% ...
\vskip\cmsinstskip
% ...
1:~~Also at Vienna University of Technology, Vienna, Austria\\
2:~~Also at CERN, European Organization for Nuclear Research, Geneva, Switzerland\\
% ...

\end{sloppypar}
\end{document}